\documentclass[12pt]{article}

\usepackage{amssymb}
\input psfig.sty
\usepackage{graphicx}
\usepackage{epsfig}
\def\be{\begin{eqnarray} &&}
\def\nonu{\nonumber \\ &&}
\def\ee{\end{eqnarray}}
\def\psla{\slash \! \! \!}
\input psfig.sty
\begin{document} 
\pagestyle{empty}
\Huge{\noindent{Istituto\\Nazionale\\Fisica\\Nucleare}}

\vspace{-3.9cm}

\Large{\rightline{Sezione di ROMA}}
\normalsize{}
\rightline{Piazzale Aldo  Moro, 2}
\rightline{I-00185 Roma, Italy}

\vspace{0.65cm}

\rightline{INFN-1363/03}
\rightline{November 2003}

\vspace{1.5cm}

\begin{center}{\large\bf Electromagnetic form factor of the pion in 
the space- and time-like 
regions within the front-form dynamics}
  
\vspace{1.cm}
 {J. P. B. C. de Melo$^a$, T. Frederico$^b$, E. Pace$^c$ and 
G. Salm\`e$^d$}
\end{center}

\noindent { $^a$ \it Instituto de F\'\i sica Te\'orica, Universidade Estadual Paulista 
0145-900, S\~ao Paulo, SP, Brazil}

\noindent { $^b$ \it Dep. de F\'\i sica, Instituto Tecnol\'ogico da Aeron\'autica, 
Centro T\'ecnico Aeroespacial, 12.228-900, S\~ao Jos\'e dos Campos,
S\~ao Paulo, Brazil}

\noindent {$^c$ \it Dipartimento di Fisica, Universit\`a di Roma "Tor Vergata" and Istituto
Nazionale di Fisica Nucleare, Sezione Tor Vergata, Via della Ricerca 
Scientifica 1, I-00133   Roma, Italy }

\noindent { $^d$ \it Istituto
Nazionale di Fisica Nucleare, Sezione Roma I, P.le A. Moro 2,
 I-00185   Roma, Italy }

\begin{abstract} 
The pion electromagnetic form factor is calculated in the space- and time-like 
regions from -10 $(GeV/c)^2$ up to 10 $(GeV/c)^2$, within 
 a front-form model. The dressed photon vertex 
where a photon decays in a quark-antiquark pair is depicted
 generalizing the vector meson dominance ansatz, 
by means of the vector meson vertex functions. 
An important feature of our model is the description of the  
on-mass-shell vertex functions in the valence sector, for the pion 
and the vector mesons, through   
the front-form wave functions
 obtained within a realistic quark model. The 
theoretical results show an excellent agreement with the data in the space-like 
 region, while in the time-like region the description is quite encouraging. 
\end{abstract} 
\vspace{2.8cm}
\hrule width5cm
\vspace{.2cm}
\noindent{\normalsize{To appear in {\bf Phys. Lett. B} }}

\newpage
\pagestyle{plain} 
 
In the framework of the front-form dynamics  
\cite{dirac} (see, e.g. \cite{kp,brodsky} for extensive reviews) 
 a large number of papers has been devoted to the study  
of 
 the electromagnetic form factor of the pion, mostly dealing with the 
 space-like (SL) region 
\cite{chung,tob92,salme,pion99,ba01,pach02,Hw}. To our knowledge, in 
the front-form quantization 
the elastic time-like (TL) form factor 
has only been calculated in a scalar field theory model  
for $q\overline Q$ mesons with point-like vertices \cite{choi01}. 
 
 In this letter, the pion electromagnetic  
form factor is evaluated within the 
front-form dynamics, in 
 both the SL and the TL regions, by using as a starting point 
the Mandelstam formula \cite{mandel} for the triangle diagrams of Figs. 1 
and 2. 
Our aim is to investigate the possibility of describing  
the photon-hadron interaction by applying the   
vector meson (VM) dominance ansatz (see, e.g., \cite{Conn})
 at the level of the photon vertex function. 
    
 The virtual processes where a quark absorbs or  
radiates a pion are present in both the SL and the TL regions
(see the square blobs in Figs. 1(b) and 2(a), 2(b), respectively) \cite{Hw}.  
In a recent study of decay processes within the 
front-form dynamics \cite{JI01} the amplitude for the pion emission  
from a quark was described by a pseudoscalar coupling of  quark and pion 
fields,  
multiplied by a constant. Here, we just follow this suggestion, and use  
a constant to parametrize  
 the amplitudes for the
radiative pion absorption or emission by a quark. 
 
In order to simplify our calculations, we evaluate  
the pion form factor for a vanishing pion mass, i.e. at the chiral limit. 
 As a consequence, the gap  
between the SL region and the TL one (i.e. between 
$q^2~\le~0$ and $q^2~\ge~4 m^2_\pi$) disappears. 
 
Our starting point is the Mandelstam covariant expression \cite{mandel} 
of the amplitudes for the processes 
 $\pi~\gamma^* \rightarrow \pi'$, or 
$\gamma^* \rightarrow \pi \pi'$, where the meson $\pi'$ is a pion 
in the elastic case or an antipion in the production process. For the TL region  
one has (see Fig. 2)  
\be 
j^{\mu} ~=~ ~-\imath e ~2  \frac{m^2}{f^2_\pi} N_c\int 
\frac{d^4k}{(2\pi)^4} 
\overline{\Lambda}_{\pi'}(k-P_{\pi},P_{\pi^{\prime}}) \overline 
\Lambda_{\pi}(k,P_{\pi}) ~\times   
\nonu  
Tr[S(k-P_{\pi}) \gamma^5 
S(k-q)~\Gamma^\mu(k,q)~S(k) \gamma^5 ] \ ,  
\label{jmu}  
\ee  
where 
$N_c=3$ is the number of colors; $\displaystyle 
S(p)=\frac{1}{\psla p-m+\imath \epsilon} \,$,  
with $m$ the mass of the constituent quark; 
$q^{\mu}$ is the virtual photon momentum; $\overline 
\Lambda_{\pi}(k,P_{\pi})$  the vertex function for the 
pion, which will be assumed to be a symmetric function of the two quark
 momenta; $P^{\mu}_{\pi}$ and 
 $P^{\mu}_{\pi^{\prime}}$ are the pion momenta. 
 The factor 2 stems from isospin algebra. 
The "bar" notation on the vertex function means that the associated 
amplitude is the solution of the Bethe-Salpeter equation where the 
two-body irreducible kernel is placed on the right of the 
amplitude, while in the conventional case it is placed on the left 
of the Bethe-Salpeter amplitude \cite{lurie}.  For the 
SL region $P^\mu_\pi$ should be replaced by $-P^\mu_\pi$, 
and then the initial pion vertex should be written as  
$\Lambda_{\pi}(-k,P_{\pi})=\overline \Lambda_{\pi}(k,-P_{\pi})$. 
 
The central assumption of the paper is our  
microscopical description of the dressed photon vertex, 
$\Gamma^\mu(k,q)$, 
in the processes where a photon with $q^+>0$ decays in a quark-antiquark pair  
at equal light-front times. In these processes 
 we approximate the plus component of the  photon vertex,  
 dressed by the interaction between the $q\overline q$ pair, as follows 
(see Fig. 3) 
\begin{eqnarray}  
\Gamma^+(k,q) &=& \sqrt{2} \sum_{n, \lambda} 
\left [ \epsilon_{\lambda} \cdot \widehat{V}_{n}(k,k-q)  \right ]    
\Lambda_{n}(k,P_n) ~ { [\epsilon ^{+}_{\lambda}]^* f_{Vn} \over \left [ q^2 - 
M^2_n + \imath M_n \Gamma_n(q^2)\right ]} \ , 
\label{cur5}  
\end{eqnarray}  
where $f_{Vn}$ is the decay constant of the $n-th$ vector 
meson in a virtual photon (see below), $M_n$  the corresponding 
 mass,  $P_n$ the VM total momentum  
($P^{\mu}_n \equiv \{P^-_n=(|{\bf q}_{\perp}|^2+M^2_n)/q^+, 
{\bf P}_{n \perp}={\bf q}_{\perp}, P^+_{n}=q^+ \}$; note that 
at the production vertex, see Fig. 3, the front-form 
three-momentum is conserved), and $\epsilon_{\lambda}$ 
the VM polarization. 
The total decay width in the denominator is assumed to be vanishing in the SL
 region, while it is 
equal to $\Gamma_n(q^2) = \Gamma_{n} ~ q^2 / M_{n}^2$ in the TL one \cite{Saku}. 
For a detailed discussion of Eq. (\ref{cur5}) see Ref. \cite{tob}. 
In Eq. (\ref{cur5}) the sum runs over  
all the possible vector mesons, and the quantity 
$\left [ \epsilon_{\lambda} \cdot \widehat{V}_{n}(k,k-q)  \right ]  ~  
\Lambda_{n}(k,q)$ is the VM vertex function.  
The momentum component, $\Lambda_{n}(k,P_n)$, 
 of the VM vertex function,   
evaluated on the quark mass shell (i.e., for $k^- = k^-_{on} =
 (|{\bf k}_{\perp}|^2 + m^2)/k^+ $),   
  will be related for $0 < k^+ < P^+_n$ to the momentum part of the  
front-form VM wave function \cite{Jaus90}, which describes the 
 valence component of 
the meson state $|n\rangle$, 
\begin{eqnarray}  
\psi_{n}(k^+, {\bf k}_{\perp}; P^+_{n}, {\bf P}_{n \perp}) =   
\frac{P^+_{n}}{[M^2_n - M^2_0(k^+, {\bf k}_{\perp}; P^+_{n}, {\bf P}_{n \perp})]} ~  
[ \Lambda_{n}(k,P_{n}) ]_{[k^- = k^-_{on}]}  ~~ ~~ .  
\label{wfn} 
\end{eqnarray} 
In Eq. (\ref{wfn}), 
$M_{0}(k^+, {\bf k}_{\perp}; P^+, {\bf P}_{\perp} )$ is the free mass 
of a $q\bar{q}$ system with total momentum $P$, and 
 individual kinematical momenta $(k^+, {\bf k}_{\perp})$ and $(P^+ - k^+, {\bf P}_\perp 
-{\bf k}_{\perp})$. 
Since in our model we consider for the moment only the $^3S_1$ vector mesons,
 we adopt, for the on-the-mass-shell spinorial part of the VM vertex, the form 
 given in  
Ref. \cite{Jaus90} (that generates the well-known Melosh rotations for $^3S_1$ 
states) 
\be 
\widehat{V}^{\mu}_{n}(k,k-q) = \gamma^{\mu} - 
{k^{\mu}_{on}-(q-k)_{on}^{\mu} \over M_{0}  
(k^+, {\bf k}_{\perp}; q^+, {\bf q}_{\perp}) + 2 m } ~~\ . 
\label{gams1}  
\ee  
 The coupling constant, $f_{Vn}$, of the $n-th$ vector meson is defined by the  
 covariant expression \cite{Jaus99} 
$\epsilon ^{\mu}_{\lambda} \sqrt{2} f_{Vn} = \langle 0| \bar{q} \gamma^{\mu} q| 
\phi_{n,\lambda}\rangle $, 
with $|\phi_{n,\lambda}\rangle$  the VM state. The coupling constant 
can be obtained from the front-form VM wave function by evaluating   
this expression with $\mu~=~+$ and $\lambda = z$,  
in the rest frame of the resonance, where  
$P_n^{\mu} = (M_n,{\vec 0} )$, $P_n^+ = M_n$
 (see Fig. 3). Assuming that $\Lambda_n(k,P_n)$ 
does not diverge in the complex plane $k^-$ for 
$|k^-|\rightarrow\infty$, and 
 neglecting the contributions of its singularities   
 in the $k^-$ integration, one obtains 
\be 
f_{Vn} = -\imath \frac{N_c} {4 (2\pi)^4} ~ 
\int dk^- dk^+ d{\bf k}_{\perp} 
\frac{Tr[ \gamma^+ ~ {\mathcal {V}}_{nz}(k,k-P_n)] ~ \Lambda_{n}(k,P_n)} 
{[k^2 - m^2 + \imath \epsilon] [(P_n - k)^2 - m^2 + \imath \epsilon]} ~ =    
\nonumber \\ 
\nonu
= - {N_c \over 4 (2\pi)^3} \int_0^{P_n^+} dk^+ ~ d{\bf k}_{\perp} 
{ Tr \left [  \gamma^+  ~  {\mathcal V}_{nz}(k,k-P_n) \right ]
\over k^+ ~ (P_n^+ - k^+)} ~ \psi_{n}(k^+, {\bf k}_{\perp}; M_{n}, {\vec 0}_{\perp}) 
 \ , 
\label{fV1}  
\ee  
where ${\mathcal V}_{nz}(k,k-P_n) = (\psla k - \psla P_n + m)  
\widehat{V}_{n z}(k,k-P_n) (\psla k + m)$.
Since both the quarks are on shell in the front-form wave function  
and there is symmetry with respect to the two quark momenta, 
we do not distinguish between $[ \Lambda_{n}(k,P_{n}) ]_{[k^- = k^-_{on}]} $ and 
$[ \Lambda_{n}(k,P_{n}) ]_{[k^- =P^-_n - (P_n-k)^-_{on}]}$ 
in the range $0 < k^+ < P^+_n$. With our assumptions, one obtains the same expression for 
$f_{Vn}$, either if the $k^-$ integration is performed in the upper or in the lower  
complex $k^-$ semiplane. 
 
For a unified description of TL and SL form factors,  
it is necessary to choose 
a reference frame where the plus component of the momentum transfer, $q^+$, 
is different from zero (otherwise, $q^2=q^+ q^- - q^2_{\perp}$ cannot be positive). 
 It is well known that 
the choice of the reference frame has a fundamental role, as shown in previous 
works in the SL region \cite{pach98,pion99,ba01,pach02} 
and in the TL one \cite{choi01}.  
In Ref. \cite{LPS} it was shown that, within the front-form 
Hamiltonian dynamics, a Poincar\'e covariant and conserved current  
operator can be obtained 
from the matrix elements of the free current, evaluated in the Breit  
reference frame,  
where the initial and the final total momenta of the system are directed  
along the 
spin quantization axis, $z$. Following Ref. \cite{LPS}, we calculate Eq. 
 (\ref{jmu}) in a 
reference frame where ${\bf q}_{\perp}=0$ and $q^+>0$. 
 
Both in the SL and the TL regions, the integration on $k^-$ in  Eq. (\ref{jmu}) is 
performed with the assumption that: i) $\Lambda_{\pi}(k,P_{\pi})$ 
does not diverge in the complex plane $k^-$ for 
$|k^-|\rightarrow\infty$, and ii) the contributions of the possible 
singularities of $\Lambda_{\pi}(k,P_{\pi})$ can be neglected. 
 Also the contributions of the poles in $k^-$ of the photon vertex 
function, $\Gamma^+(k,q)$, are supposed to be negligible. 
  
Then, in the SL case, where $P^{\mu}_{\pi^{\prime}} = 
P^{\mu}_{\pi} + q^{\mu}$, the current matrix element 
$j^{\mu}$ becomes the sum of two contributions, 
corresponding to the diagrams of Figs. 1(a) and 1(b), respectively,  
$j^{\mu}_{SL} = j^{(I) \mu}_{SL} + j^{(II) \mu}_{SL} $ \cite{pach02}.  
The contribution $j^{(I) \mu}_{SL}$ has the integration on $k^+$ constrained by 
$-P^+_{\pi} \le k^+ \le 0$, and $j^{(II) \mu}$ has the 
integration on $k^+$ in the interval $0 < k^+ < q^+$. The valence 
component of the pion contributes to  $j^{(I) \mu}_{SL}$ only, while 
 $j^{(II) \mu}_{SL}$ is the contribution 
of the pair-production mechanism from an incoming virtual photon 
with $q^+ \ > \ 0$ \cite{sawicki,pach98,pion99,ba01,pach02,ba02}. 
In the SL case we adopt a frame where   
${\bf P}_{\pi \perp} = {\bf P}_{\pi \prime \perp} = {\bf 0} $, 
and we obtain
${P}^+_{\pi} = q^+ (-1 + 
\sqrt{1-4m^2_\pi/ q^2})/2 $.  
Then, in the limit $m_\pi \rightarrow 0$ 
the longitudinal momenta of the pions are 
${P}^+_\pi=0  \ \ {\rm{and}} \ \ {P}^+_{\pi \prime} = q^+$. 
Therefore  only the contribution 
of the pair-production mechanism, $j^{(II) \mu}_{SL}$, survives (Fig. 1(b)).  
As shown in Ref. \cite{pach02}, in a frame where $q^+ > 0$,  
$j^{(II)\mu}_{SL}(q^2)$ dominates the form 
factor at high momentum transfer. Moreover, it turns out that  
 in the model of Ref. \cite{pach02} the momentum region, where  
 $j^{(II)\mu}_{SL}(q^2)$  
starts to dominate the form factor, tends toward 
zero if the pion 
mass is artificially decreased, in agreement with our present discussion.  
  
In the TL case, one has $P^{\mu}_{\pi^{\prime}} \equiv 
P^{\mu}_{\bar{\pi}}$, $q^{\mu} = P^{\mu}_{\pi} + 
P^{\mu}_{\bar{\pi}}$.  
  The integration range on $k^+$ for the matrix element of the  
current, $j^{\mu}$, can be 
decomposed in two intervals, $0 < k^+ < P^+_{\pi}$ and $P^+_{\pi} 
< k^+ < q^+$, and then $j^{\mu}_{TL}$ becomes the sum of two contributions, 
corresponding to different $x^+$-time orderings (see diagrams in Figs. 2(a) and 2(b), respectively).  
 In the final state of the $\pi \bar{\pi}$  pair we make the purely longitudinal choice,  
${\bf P}_{\bar{\pi} \perp} = -{\bf P}_{{\pi} \perp}={\bf 0} $.  Then, one obtains 
$P^+_{{\pi}}/q^+ = x_\pi = (1\pm\sqrt{1-4m^2_\pi/ q^2})2$. 
In the limit $m_\pi \rightarrow 0$, one has $x_\pi=1$ or $ 0$.  
 Analogously to the SL case, in what follows we adopt the choice $x_\pi=0$, 
which implies $P^+_\pi=0$ and $P^+_{\bar{\pi}}=q^+$. Therefore only the contribution corresponding to the  
diagram of Fig. 2(b) survives. 
 
The form factor of the pion in the 
TL and in the SL regions can be obtained from the plus 
component of the proper current matrix elements: 
$j^{\mu}_{TL} =\langle \pi \bar{\pi}| \bar{q} \gamma^{\mu}q 
|0\rangle = \left (P^{\mu}_{\pi} -P^{\mu}_{\bar{\pi}} \right )~F_{\pi}(q^2)$ , 
and
$j^{\mu}_{SL} =\langle \pi | \bar{q} \gamma^{\mu}q |\pi ^{\prime}\rangle =  
\left (P^{\mu}_{\pi} + P^{\mu}_{\pi ^{\prime}} \right )~F_{\pi}(q^2)$  
Since in the limit $m_\pi \rightarrow 0$ the form factor 
 receives contributions only from the diagrams of Figs. 1(b) and 2(b),  
where the photon decays in a $q\overline q$ pair,  
one can apply our approximation  for the plus component 
of the dressed photon vertex (\ref{cur5}), 
both in the SL and in the TL regions. 
Then the matrix element $j^+$ can be written as a sum  
 over the vector mesons and 
 consequently the form factor becomes 
 \be 
 F_{\pi}(q^2) = \sum_n~ {f_{Vn} \over q^2 - M^2_n + \imath M_n \Gamma_n(q^2)} ~ 
 g^+_{Vn}(q^2)~~ ~~ , 
\label{tlff}  
\ee 
where $g^+_{Vn}(q^2)$, for $q^2 > 0$, is the form factor for the VM decay in a pair of pions. 
 
Each VM contribution to the sum (\ref{tlff}) is 
 invariant under kinematical front-form boosts and 
therefore 
it can be evaluated in the rest frame of the  
corresponding resonance. 
 In this frame one has $q^+=M_n$ and $q^-=q^2/M_n$ for the photon and 
$P^{+}_n= P^{-}_n = M_n$  for the vector meson.   
  This means that we choose a different frame for 
each resonance (always with ${\bf q}_{\perp}=0$), 
 but all the frames are related by kinematical front-form boosts 
along the $z$ axis to each other, and to the frame where $q^+ = -q^- = \sqrt{-q^2}$ 
~ ($q_z=\sqrt{-q^2}$), adopted  in previous analyses of the SL region 
\cite{LPS,pach02}. 
Since in our reference frame one has 
$\sum_{\lambda} \left [ \epsilon 
^{+}_{\lambda}(P_n) \right ]^* \epsilon _{\lambda}(P_n) \cdot 
\widehat{\Gamma}_{n} = \left [ \epsilon ^{+}_{z}(P_n) \right ]^* 
\epsilon _{z}(P_n) \cdot \widehat{\Gamma}_{n} = - 
\widehat{\Gamma}_{n z}$, 
 we obtain from Eqs. (\ref{jmu},\ref{cur5},\ref{wfn}) 
\be  
g^+_{Vn}(q^2) =  {N_c \over 8 \pi^3} { \sqrt{2} \over P^+_{\overline{\pi}} }  
\frac{m}{f_\pi} ~ \int_0^{q^+} {dk^+ \over (k^+)^2~(q^+-k^+)} \int d{\bf k}_{\perp} ~  
Tr \left [ [\Theta^z ~ 
\overline{\Lambda}_{\pi}(k ; P_{\pi})]_{(k^-=q^- + (k - q)^-_{on})} \right ]  
\times 
\nonu  
\nonu   
\psi^*_{\overline{\pi}}(k^+, {\bf k}_{\perp}; 
P^+_{\overline{\pi}},{\bf P}_{\overline{\pi} \perp })  
~{[M_n^2 - M^2_0(k^+, {\bf k}_{\perp}; q^+, {\bf q}_{\perp})]  
\over [q^2-M^2_0(k^+, {\bf k}_{\perp}; q^+, {\bf q}_{\perp})+i\epsilon]}  
\psi_{n}(k^+, {\bf k}_{\perp}; q^+,{\bf q}_{\perp}) ~  \ ,  
\label{ffpi2} 
\ee  
where  $\Theta^z = {\mathcal V}_{nz}(k,k - q) 
~\gamma^5~ \left [\psla k - \psla P_{\pi} + m \right ] ~ \gamma^5$.   
To obtain Eq. (\ref{ffpi2}) we have first  
performed the $k^-$ integration, and then we have related 
$[ \overline{\Lambda}_{\overline{\pi}}(k - P_{\pi},P_{\overline{\pi}}) ] 
_{(k^- = q^- + (k - q)^-_{on})}$ in the valence sector
 to the momentum component of the corresponding front-form pion wave function 
 through the following equation (see Eq. (\ref{wfn})) 
\be  
\psi_{\pi}(k^+, {\bf k}_{\perp}; P^+_{\pi}, {\bf 
P}_{\pi \perp}) = ~ \frac{m}{f_\pi} ~ 
\frac{P^+_{\pi}~ [ \Lambda_{\pi}(k,P_{\pi}) ]_{[k^- = k^-_{on}]} }
{[m^2_\pi - M^2_0(k^+, {\bf k}_{\perp}; P^+_{\pi}, {\bf P}_{\pi \perp})]} ~~ ~~ .  
\label{wf2} 
\ee 
As for the VM vertex function, 
we do not distinguish between 
$~[ \Lambda_{\pi}(k,P_{\pi}) ]_{[k^- = k^-_{on}]}~ $ and 
$~[ \Lambda_{\pi}(k,P_{\pi}) ]_{[k^- =P^-_\pi - (P_\pi-k)^-_{on}]}~$, 
in the range $0 < k^+ <P^+_\pi$. 
 
 Following Ref. \cite{JI01}, in our model 
 the momentum part of the quark-pion emission  vertex in the non-valence sector
which appears in Eq. (\ref{ffpi2}), 
$  \frac{m}{f_\pi} 
[\overline{\Lambda} _{\pi}(k ; P_{\pi})]_{(k^-=q^- + (k - q)^-_{on})} $  
(see the square blob in Fig.  2(b)), is assumed to be a constant. 
The value of the constant is fixed by the pion charge normalization. 
 The same constant value is assumed for  
 the quark-pion absorption vertex (see the square blob in Fig. 1(b)). 
 
Let us stress that, within our assumptions, 
$g^+_{Vn}(q^2)$ is given by the same expression 
 both in the TL and in the SL 
 ($P_{\pi}~\rightarrow~-P_{\pi}$, see below Eq. (\ref{jmu}))  
 regions. Indeed 
 in the limit $m_\pi \rightarrow 0$, one has $P^+_{\pi} = 0$,  
 ${\bf P}_{\pi\perp} = 0$, and $\pm ~P^-_{\pi} = q^2/M_n$ with the positive
  (negative)  
 sign in the TL (SL) region, and therefore the sign in front of  
 $\psla P_{\pi}$ in the quantity $\Theta^z$ of Eq. (\ref{ffpi2}) has no effect. 
 Then 
the  form factor is continuous in this limit, at $q^2=0$. 
  It turns out that for $m_\pi=0$ only the instantaneous terms (in $x^+$-time,
   see, e.g., \cite{pach02})
  contribute to Eq. (\ref{ffpi2}) \cite{tob}. 
 
 In order to describe the pion and the interacting $q\bar q$ pairs in the
  $1^-$ channel, 
  we use the eigenfunctions of a square 
mass operator proposed in Refs. \cite{tobpauli,FPZ02}, within a relativistic  
 constituent quark 
model. This model  takes into account  
confinement through a harmonic oscillator potential and the  
$\pi-\rho$ splitting through a Dirac-delta interaction in the pseudoscalar  
channel. It achieves a  
satisfactory description of the experimental masses 
for both singlet and triplet $S$-wave mesons, with a  natural 
explanation of the "Iachello-Anisovitch law"~\cite{Iach,ani}, namely the almost-linear  
relation 
between the square mass  of the excited states and the radial quantum 
number $n$. Since the model of Refs. \cite{tobpauli,FPZ02}  
does not include the mixing between isoscalar and 
isovector mesons, in this paper we include only the contributions of the isovector 
$\rho$-like vector mesons. 
 
The eigenfunction $\psi_{n}(k^+, {\bf k}_{\perp}; q^+,{\bf q}_{\perp})$,   
 which describes  the valence component of 
the meson state $|n\rangle$, is normalized to the probability 
of the lowest ($q\bar q$) Fock state (i.e. of the valence component). The $q\bar q$ 
probability can be roughly estimated to be $\sim 1/\sqrt{2n + 3/2}$ in a 
simple model \cite{tob} 
that reproduces the "Iachello-Anisovitch law" \cite{Iach,ani}, 
and is based on an expansion of the  
VM state $|n\rangle$, in terms of properly weighted 
Fock states $|i\rangle_0$, 
with $i > 0$ quark-antiquark pairs.

Our calculation of the pion form factor contains a very small set of parameters:  
i) the constituent quark mass, ii)  the 
oscillator strength, $\omega$, and iii) the VM 
 widths, $\Gamma_n$, for $M_n >~2.150~GeV$. The up-down quark mass is fixed at 
0.265 $GeV$ \cite{FPZ02}. 
For the first four vector mesons  
the known experimental masses and widths are used in the calculations \cite{pdg}. 
However, the non-trivial $q^2$ dependence of $g^+_{Vn}(q^2)$ in our microscopical model 
implies a shift of the VM masses, with respect to the values  
obtained by using  
Breit-Wigner functions with constant values for $g^+_{Vn}$.  
As a consequence,  
 the value of the $\rho$ meson mass is 
moved in our model from the usual one, $.775 ~ GeV$, to $.750~GeV$. 
For the other VM, the mass values corresponding to the model of Ref. \cite{FPZ02} 
are used, while for the unknown widths we use a single value 
$\Gamma_n = 0.15 ~GeV$, which presents the best agreement with the compilation of
the experimental data of 
Ref. \cite{baldini}. We consider 20 resonances in our calculations 
to obtain  stability of the results up to $q^2 =~10 ~(GeV/c)^2$.

The oscillator strength is fixed at $\omega = 1.39~GeV^2$  \cite{ani} .  
 The values of the coupling constants, $f_{Vn}$, are evaluated from the model VM  
wave functions through Eq. (\ref{fV1}). The corresponding 
partial decay width \cite{Jaus99} 
$\Gamma_{e^+e-} = 8\pi \alpha^2 ~ f_{Vn}^2 / ( 3 M_n^3 )$,  
 where $\alpha$ is the fine structure constant,   
 can be considered to be in agreement with the experimental data 
for the $\rho$ meson ($\Gamma^{th}_{e^+e-}$ = 6.37 KeV,  
$\Gamma^{exp}_{e^+e-}$ = 6.77~ $\pm~ 0.32$ KeV ), and for 
$\rho^\prime$ and $\rho^{\prime\prime}$ ($\Gamma^{th}_{e^+e-}$ = 1.61 KeV 
and $\Gamma^{th}_{e^+e-}$ = 1.23 KeV, respectively) 
 to be consistent with the experimental lower 
bounds ($\Gamma^{exp}_{e^+e-} > 2.30 ~\pm 0.5  $ KeV 
and $\Gamma^{exp}_{e^+e-} > 0.18 ~\pm 0.1$ KeV, respectively) \cite{pdg}. 
 
We perform two sets of calculations. In the first one, we use the 
asymptotic form of the pion valence wave function, obtained  
with $\Lambda_{\pi}(k,P_{\pi}) = 1 $ in Eq. (\ref{wf2});  
 in the second one, we use the eigenstate of 
the square mass operator of Refs. \cite{tobpauli,FPZ02}. The pion radius found for the 
asymptotic wave function is $r_\pi^{\rm{asymp}}$ = 0.65 fm  
and for the full model wave function is $r_\pi^{\rm{model}}$ = 0.67 
fm, to be compared with the experimental value  
$r_\pi^{\rm{exp}}$ = $0.67 ~\pm ~0.02$ fm \cite{amen}. The good 
agreement with the experimental form factor at low momentum 
transfers is expected, since we have built-in 
the generalized $\rho$-meson dominance. 
 
The calculated pion form factor is shown in Fig. 4 
in a wide region of square 
momentum transfers, from $-10$ $(GeV/c)^2$ up to 10 $(GeV/c)^2$. A general 
qualitative agreement with the data is seen, independently of the detailed form of the pion 
wave function.  
The results obtained with the asymptotic pion wave function and the full model, present 
some differences only above $3 ~(GeV/c)^2$. 
The SL form factor is notably 
well described, except near -10 $(GeV/c)^2$.
It has to be stressed that the heights of the TL bumps directly depend 
on the calculated values of $f_{Vn}$ and $g^+_{Vn}$. 
   
The introduction of $\omega$-like \cite{Gard} 
and $\phi$-like mesons  
could  improve the description of the data in the TL region.
For instance, the introduction of these mesons could smooth out the oscillations of the form factor
at high momentum 
transfer values.
 However, a consistent dynamical description 
 of $\omega$-like
and $\phi$-like states is far beyond the present work, and we leave it for 
 future developments of the model.
 
Our results show that a VM dominance ansatz for the (dressed photon )-($q\overline q$)
vertex, within a model consistent with the meson spectrum, 
is able to give a unified description of the SL and TL pion form factor. Using the 
 experimental widths for the first four vector mesons and a single free 
 parameter for the unknown widths of the other vector mesons, the model gives 
 a qualitative agreement with the TL data, while in the SL region it works surprisingly well. 
 Our VM dominance model can be also applied to evaluate other observables, 
 as the $\gamma^* \rightarrow \pi ~ \gamma$ form factor or the nucleon TL form factor.

\section*{Acknowledgments} 
This work was partially supported by the Brazilian agencies CNPq 
and FAPESP and by Ministero della Ricerca Scientifica e 
Tecnologica. J.P.B.C.M. and T.F. acknowledge the hospitality of 
the Dipartimento di Fisica, Universit\`a di Roma "Tor Vergata" and 
of Istituto Nazionale di Fisica Nucleare, Sezione Tor Vergata and 
 Sezione Roma I.

 

\newpage 
 
\begin{figure}[t] 
\includegraphics[width=15.cm]{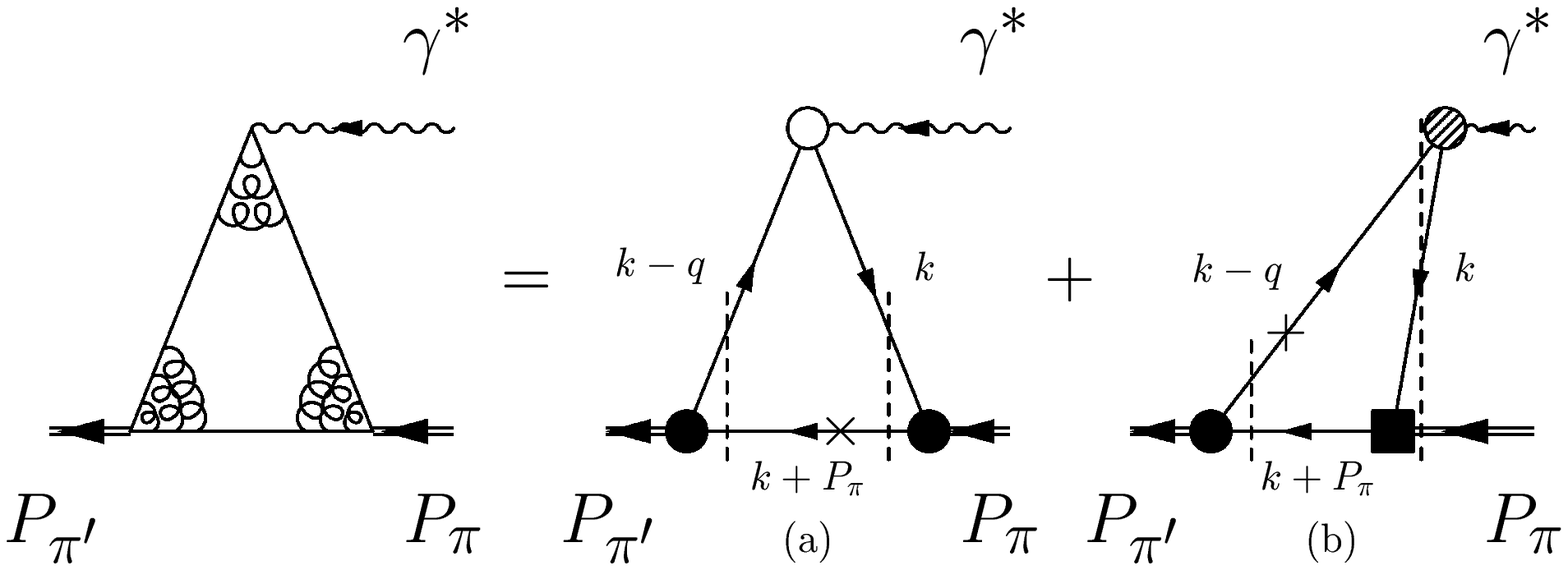}
 
\vspace{-13cm}
Figure 1. 
Diagrammatic representation of the pion 
elastic form factor for  $q^+ >0$ vs the global $x^+$-time flow. 
 Diagram $(a)$ 
($0\le -k^+ \le P^+_{\pi}$) is the contribution of the 
valence component in the initial pion wave function. 
Diagram $(b)$ ($P^+_{\pi} \le -k^+ \le {P'}^+_{\pi}$) is 
the non-valence contribution to the pion form factor. Both 
processes contain the contribution from the dressed photon 
vertex. The crosses correspond to the  quarks on the $k^-$ shell (see text).
\end{figure}

\begin{figure}[t]
\includegraphics[width=15.cm]{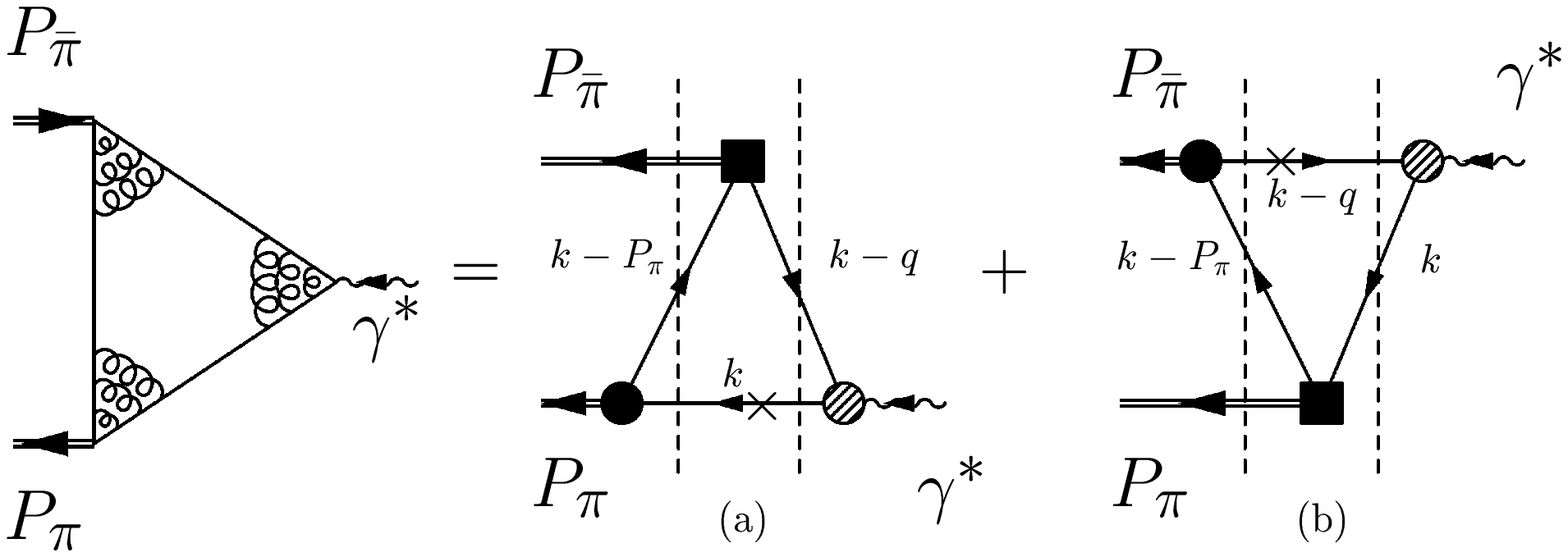}

\vspace{-13cm}
Figure 2. 
Diagrammatic representation of the photon decay  
($\gamma ^* \rightarrow \pi \bar{\pi}$) vs the global $x^+$-time flow. 
Diagrams (a) and (b) correspond to different  $x^+$-time orderings. 
The crosses correspond to the  quarks on the $k^-$ shell (see text). 
\end{figure} 
 

\begin{figure}[t] 
\includegraphics[width=15.cm]{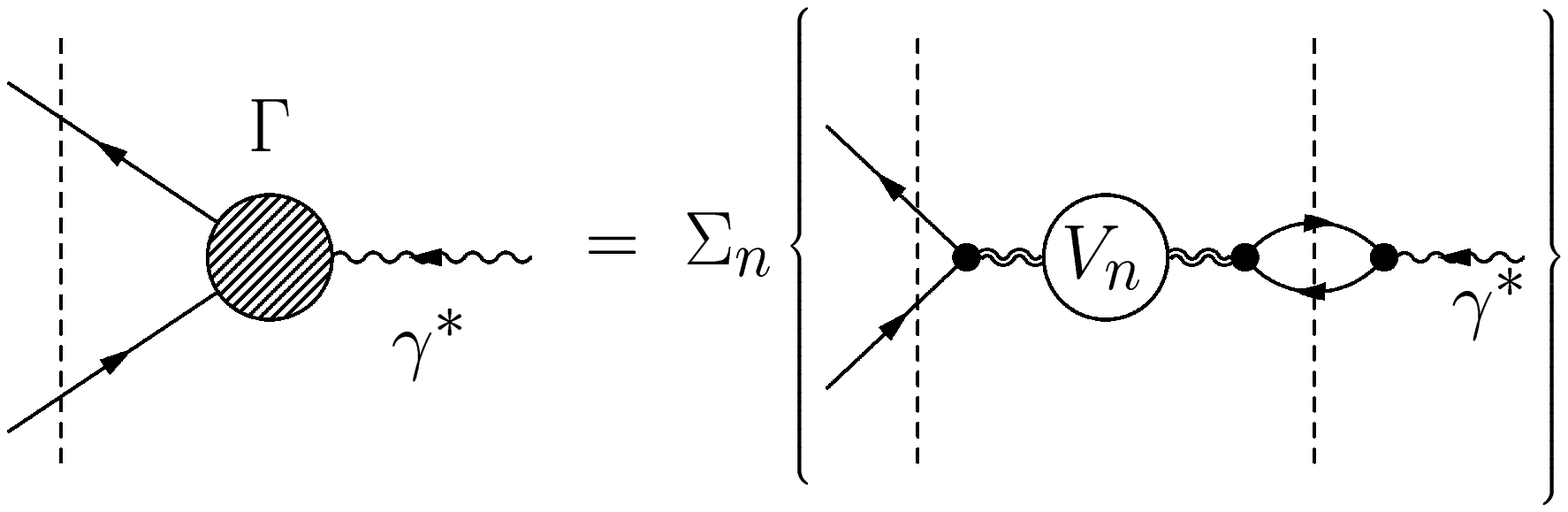}

\vspace{-13cm}
Figure 3. 
Dressed photon vertex. The double-wiggly 
lines represent the front-form Green function describing the 
vector-meson propagation.  
\end{figure} 

\begin{figure}[t] 
\hspace{-1.cm}\includegraphics[width=16cm]{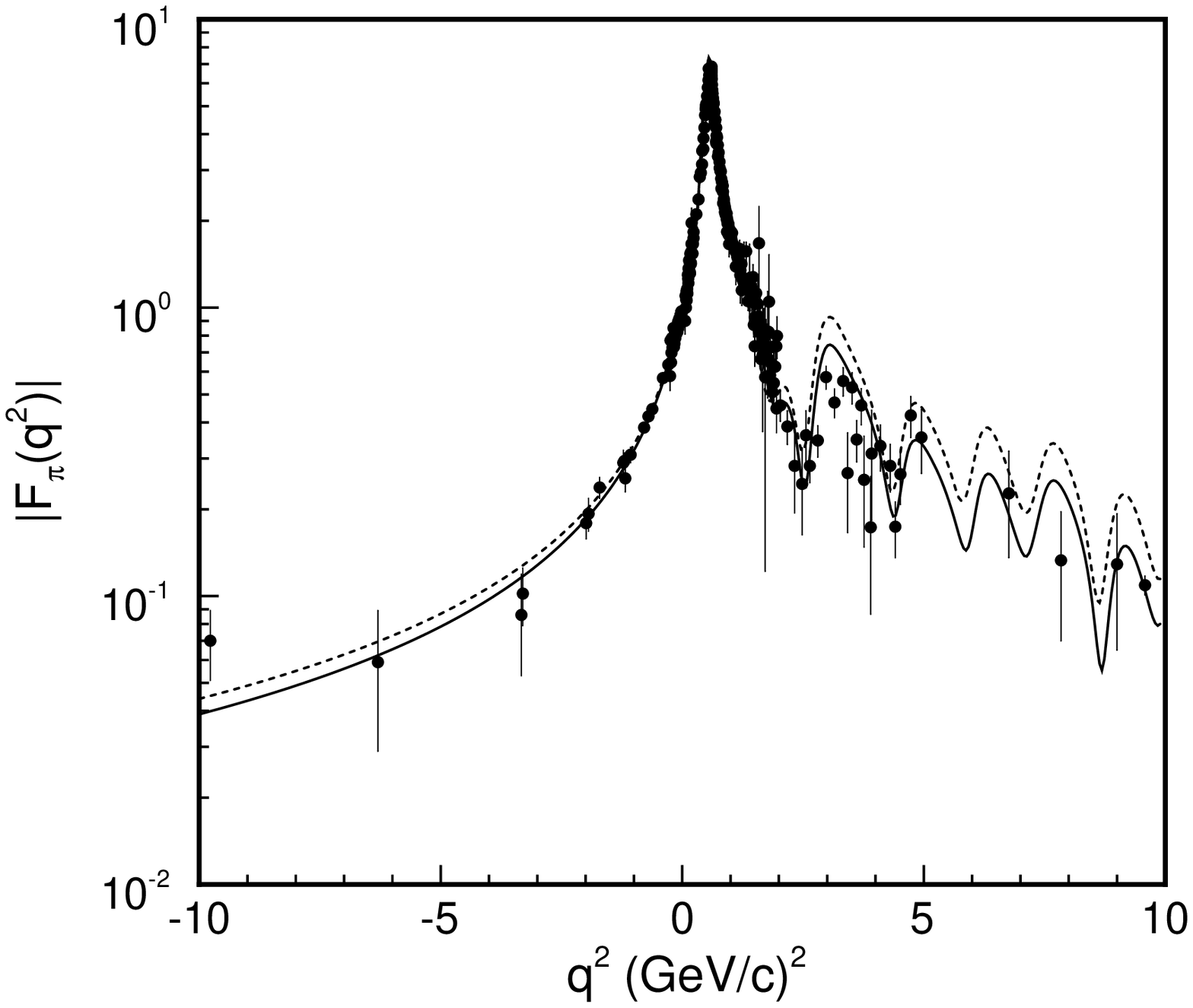} 

\vspace{1cm}
Figure 4. 
Pion electromagnetic form factor vs the 
square momentum transfer $q^2$. Dashed and solid lines 
are the results with 
 the asymptotic (see text) and the full pion wave function, respectively. 
Experimental data are from Ref. \cite{baldini}. 
\end{figure}

\end{document}